\documentclass[prl,twocolumn,amssymb,amsmath]{revtex4-1}
\usepackage{graphicx}
\usepackage{epsfig,color}
\usepackage{amsmath,amssymb}
\begin{document}

\title{String tension and area-law probed using quantum superposition}

\date{\today}

\author{Erez Zohar}
\affiliation{School of Physics and Astronomy, Raymond and Beverly Sackler
Faculty of Exact Sciences, Tel-Aviv University, Tel-Aviv 69978, Israel.}
\author{Benni Reznik}
\affiliation{School of Physics and Astronomy, Raymond and Beverly Sackler
Faculty of Exact Sciences, Tel-Aviv University, Tel-Aviv 69978, Israel.}

\begin{abstract}
We propose a method for measuring the string tension in gauge theories,
by considering an interference effect of mesons,  which is governed by a space-time area law, due to
confinement.  Although it is only a gedanken experiment for real elementary particles, in the context of quantum simulations of confining gauge theories such an experiment can be realized.
\end{abstract}

\maketitle

Free quarks can not be found in nature, but rather form hadrons, due to the phenomenon of quark confinement \cite{Wilson}. A quark and an anti-quark, attached to each other by a confining flux-tube, form a meson, which is the simplest hadron in QCD. The static potential between the quarks, as a function of their distance $R$, takes the form
\begin{equation}
V\left(R\right) = \gamma R
\end{equation}
for large values of $R$, where $\gamma=\gamma(g^2)$ is called the string tension, and $g$ is the coupling constant \cite{Wilson,KogutSusskind,Polyakov,KogutLattice}.

Wilson Loops manifests confinement through the area dependence. However they involve
           a product of operators along a loop and thus are non-local.
            We propose an alternative method that circumvents this difficulty and
yet is sensitive to the same area  dependence.

Consider two \emph{static} quarks, separated by distance $R$. Being static, they are entangled with the gauge field, and thus we can formulate our discussion in terms of gauge field states. Taking confinement into account, we know that the two quarks are connected by a long flux-tube, with length $R$, forming a meson (Quarkonium). We shall denote this state as $\left|R\right\rangle$. Due to the confinement, the energy of this state satisfies
\begin{equation}
H \left|R\right\rangle =  \gamma R \left|R\right\rangle
\end{equation}
where $H$ is the gauge field's Hamiltonian.

Suppose that one of the quarks is then moved, along the straight line connecting the quarks, a distance $L$ further apart from its original position, using a quark translation operator, denoted by $\mathcal{T}$. Due to Gauss's law and gauge invariance, the flux tube will "follow" the quark to its new position (as a manifestation of the entanglement between the fermionic and gauge field degrees of freedon), creating, eventually, a state $\mathcal{T}\left|R\right\rangle=\left|R+L\right\rangle$, satisfying
\begin{equation}
H \left|R+L\right\rangle =  \gamma \left( R + L \right) \left|R+L\right\rangle
\end{equation}

In the subspace of these two states, we can reformulate our system as an effective two-level system, with
\begin{equation}
\left|\uparrow\right\rangle \equiv \left|R+L\right\rangle  ;
\left|\downarrow\right\rangle \equiv \left|R\right\rangle  ;
H=\gamma R  +\gamma L\left|\uparrow\right\rangle\left\langle\uparrow\right|
\end{equation}
and identify the quark translation operator as the spin raising operator $\sigma_{+}$.

Thus, using a unitary transformation $U=e^{i \frac{\pi}{4} \sigma_{y}}$ one could generate a state which is a superposition of two mesonic states, with two different string lengths:
\begin{equation}
\left|\psi\right\rangle
=U\left|\downarrow\right\rangle
=\frac{1}{\sqrt{2}}\left(\left|\downarrow\right\rangle + \left|\uparrow\right\rangle\right)
=\left|\uparrow_{x}\right\rangle
\end{equation}

 \begin{figure}[t]
\includegraphics[scale=0.45]{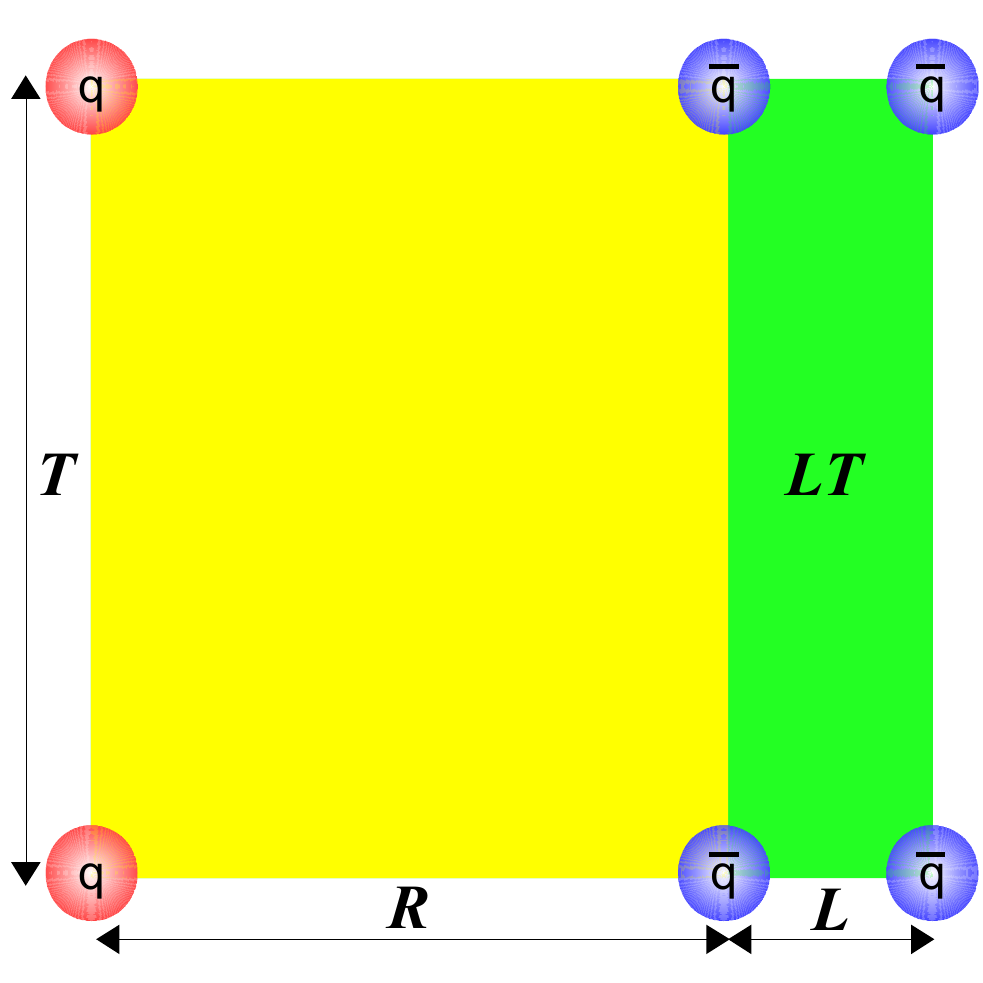}
\caption{An example of the area law. The antiquark is in a superposition of two locations, separated by a distance $L$. The phase difference between the two branches is proportional to
the area $LT$. }
\end{figure}
After a time period $T$, the state evolves to
\begin{equation}
\left|\psi\left(T\right)\right\rangle
=\frac{1}{\sqrt{2}}\left(e^{-i \gamma R T}\left|\downarrow\right\rangle + e^{-i \gamma \left(R+L\right) T}\left|\uparrow\right\rangle\right)
\end{equation}
- and a relative phase $-i \gamma L T$ develops between the states in superposition.
Next, apply $U$ on the state of the system again, to obtain
\begin{equation}
U\left|\psi\left(T\right)\right\rangle
=\frac{1}{\sqrt{2}}\left(e^{-i \gamma R T}\left|\uparrow_{x}\right\rangle + e^{-i \gamma \left(R+L\right) T}\left|\downarrow_{x}\right\rangle\right)
\end{equation}
or, after switching back to the $z$ (position) basis, up to a global phase
\begin{equation}
\cos\left(\frac{\gamma L T}{2}\right)\left|\uparrow\right\rangle
+ i\sin\left(\frac{\gamma L T}{2}\right)\left|\downarrow\right\rangle
\end{equation}

Now one has to measure the position of the quark in superposition. The probabilities of finding the "short" meson $\left|R\right\rangle$
and the "long" meson $\left|R+L\right\rangle$ are, respectively,
\begin{equation}
P_{R}=\sin^2\left(\frac{\gamma L T}{2}\right) ;
P_{R+L}=\cos^2\left(\frac{\gamma L T}{2}\right)
\label{prob}
\end{equation}
The probabilities depend on the phase, which depends on the area of the space-time path enclosed by the two superposition branches (as in the figure), and is proportional to the string tension $\gamma$.
This can be used to determine the string tension's dependence on the coupling constant $g$.
The above area depencence is expected in any abelian and non-abelian gauge theory within the confining phase.

 Let us examine the relation of the proposed method with the Wilson Loop operator.
 First, one creates an eigenstate state of heavy $Q \bar Q$, separated by distance $R$. We call this state $\left|R \left( g^2 \right) \right \rangle$, and write it in terms of the gauge
 field degrees of freedom.
 Assuming confinement, this corresponds to a quarkonium state, where the two quarks are connected by a flux tube; In particular, in the strong coupling limit one gets
 \begin{equation}
 \underset{g^2 \rightarrow \infty}{\text{lim}}\left|R\left(g^2\right)\right\rangle = e^{i \int_{0}^{R}A_{\mu}dx^{\mu}}\left| 0 \right\rangle
  \end{equation}
  For simplicity, we have used an abelian theory. The more general non-abelian calculation of a Wilson Loop includes some more mathematical details (as, for example, in \cite{PolyakovBook}). After a time period $T$, the state remains the same, but a phase is acquired. Thus, the amplitude to remain in the same state after time $T$ will be a pure phase,
  \begin{equation}
 B\left(g^2,R,T\right) \equiv \left\langle R\left(g^2\right)  \right| e^{-iHT} \left| R\left(g^2\right) \right\rangle = e^{-i \gamma RT}
 \end{equation}
 This phase is area-dependent: it depends on the area $RT$ of a space-time loop. In particular, in the strong coupling limit $g^2 \rightarrow \infty$, this equals the expectation value of the Wilson-Loop operator on the very same loop:
 which is the expectation value of a space-time Minkowskian (in the temporal gauge):
  \begin{equation}
    B\left(\infty,R,T\right)=\left\langle 0 \right| e^{iHT}e^{-i \int_{0}^{R}A_{\mu}dx^{\mu}}e^{-iHT}e^{i \int_{0}^{R}A_{\mu}dx^{\mu}}\left|0\right\rangle
 \end{equation}

This relates to our suggestion in the following way: we have considered a superposition of two quarkonium states, $\left| \psi \left(T\right) \right\rangle$. Measuring the amplitude of each of the superposition constituents at time $T$ results in the phase
\begin{equation}
B \left(g^2,\alpha,T\right) = \sqrt{2} \left\langle \alpha \left(g^2\right) \right|\psi\left(T\right) \rangle
\end{equation}
- where $\alpha = R,R+L$ - the amplitude of finding the moved quark in any of its possible locations, in which the relevant phases are global and hence nonobservable.
What we suggest, instead of measuring the probability of a specific quarkonium state in time $T$, is to apply the unitary transformation $U$ at time $T$ and then measure. This mixes the two states, causing them to interfere, transferring the relative phase of the Wilson Loops to the arguments of the probablities. The corrseponding amplitudes are thus
\begin{equation}
A \left(\alpha,T\right) =\left\langle \alpha \right| U \left| \psi\left(T\right) \right\rangle
\end{equation}
from which the probabilities $P_\alpha= \left| A\left(\alpha,T \right) \right|^2$ (see equation (\ref{prob})) are derived.

Interestingly, the two states in superposition experience \emph{different} electric fields. This is a reminiscent of the concept of "private potential" \cite{Private}.

In the Coulomb phase (as in 3+1 QED) or in any other $V \propto R^{\beta}$ phase, with $\beta \neq 1$, the appropriate gauge field state does not include a flux tube as in the confining phase. Hence, the final probabilities will not manifest a simple area dependence. This can be used to probe a transition between confining and non-confining phases.

We have presented a method to measure the string tension of a confining flux-tube using a superposition of mesons. Although it is only a gedanken experiment in the context of real-world
physics, such an experiment may be realized within a quantum simulation of confining gauge theories \cite{Zohar2011,Zohar2012}.

\emph{Acknowledgments.} The authors would like to thank S. Nussinov and B. Svetitsky for helpful discussions.
BR acknowledges the support of the Israel Science Foundation, the German-Israeli
Foundation, and the European Commission (PICC).
\bibliography{ref}

\end{document}